\documentclass[pdftex,twocolumn,epjb]{svjour3}          
\RequirePackage[T1]{fontenc}
\smartqed  
\RequirePackage{graphicx}
\RequirePackage{mathptmx}      
\RequirePackage{flushend}
\RequirePackage[numbers,sort&compress]{natbib}
\RequirePackage[colorlinks,citecolor=blue,urlcolor=blue,linkcolor=blue]{hyperref}

\journalname{arXiv}

\usepackage{dcolumn}   
\usepackage{bm}        
\usepackage{amssymb}
\usepackage{amsmath}
\usepackage{mathrsfs}
\begin{document}
	
\newcommand{\be}{\begin{equation}}
\newcommand{\ee}{\end{equation}}
\newcommand{\ba}{\begin{eqnarray}}
\newcommand{\ea}{\end{eqnarray}}

\newcommand{\ch}{{\cal H}}
\newcommand{\ce}{{\cal E}}

\title{One Dimensional Localization for Arbitrary Disorder Correlations}


\author{Hichem Eleuch \and Michael Hilke}

\institute{H. Eleuch \at Institute for Quantum Science and Engineering,
	Texas A$\&$M University, College Station, Texas 77843, USA.\\
	College of Arts and Sciences, Abu Dhabi University, Abu Dhabi, United Arab Emirates.
\email{heleuch@fulbrightmail.org}
\and M. Hilke
\at Department of Physics, McGill University, Montr\'eal, QC H3A 2T8, Canada and Center for the Physics of Materials (CPM).
\email{hilke@physics.mcgill.ca}
}


\maketitle

\begin{abstract}
We evaluate the localization length of the wave solution of a random potential characterized by an arbitrary autocorrelation function. We go beyond the Born approximation to evaluate the localization length using a non-linear approximation and calculate all the correlators needed for the localization length expression. We compare our results with numerical results for the special case, where the autocorrelation decays quadratically with distance. We look at disorder ranging from weak to strong disorder, which shows excellent agreement. For the numerical simulation,  we introduce a generic method to obtain a random potential with an arbitrary autocorrelation function. The correlated potential is obtained in terms of the convolution between a Wiener stochastic potential and a function of the correlation.

\end{abstract}

\section{Introduction}

Disordered systems are playing an important role in materials physics \cite{cutler1969observation,thouless1979ill,berthier2011theoretical}, cold atoms \cite{billy2008direct,roati2008anderson}, optical waveguides \cite{elson1983relationship,john1987strong,topolancik2007experimental,hilke2009seeing,karbasi2014image,skipetrov2014optical}, acoustic and phononic systems \cite{weaver1990anderson}, many-body systems \cite{basko2006metal,bardarson2012unbounded} and even time fluctuations \cite{fishman1982chaos,brandenberger2012towards}. While in some cases physical properties depend on a particular disorder configuration, most properties depend on their configurational average \cite{hilke2008ensemble}. A good example being the resistance through a macroscopic disordered system. If the system size is much greater than the coherence length,  the resistance can be computed by doing a configurational average \cite{huckestein1995scaling}. In this case only the statistical properties of the disorder are important, particularly the autocorrelator. There has been a long history of important results based on the assumption of uncorrelated disorder (or white disorder), in particular, the seminal result by Anderson \cite{anderson1958absence} on the localization of all states in one dimension \cite{thouless1979ill,abrahams1979scaling,kunz1980spectre}. In general, the solution to a problem with disorder is challenging, yet the assumption of uncorrelated disorder greatly simplifies the evaluation of averaged properties \cite{pendry1994symmetry}. However, in many physical systems, uncorrelated disorder is not a valid assumption, like for instance in speckle potentials \cite{cao2003lasing,sanchez2007anderson,hilke2017anderson} or smooth random potentials \cite{de1998delocalization,izrailev1999localization,shima2004localization,izrailev2012anomalous,eleuch2015localization}. In fact, correlations in the disorder can lead to delocalization in 1D \cite{erdos1982theories,flores1989transport,dunlap1990absence,flores1993absence,sanchez1994suppression,hilke1997delocalization,bellani1999experimental} and 2D \cite{hilke1994local,hilke2003noninteracting}. Hence finding tools to address systems where the disorder is not just uncorrelated but defined by some correlation function is crucial.

In some cases it is possible to use the Born approximation in order to find the disorder averaged properties, such as localization. This approach works well when coherent multiple scattering is neglected, which is often the case for weak disorder. Properties such as the mean free path or localization then simply depend on the Fourier transform of the disorder potential. For more general disorder potentials other methods have to be used, such as perturbation expansion \cite{sanchez1994suppression,tessieri2002delocalization} or phase averaging. Here we discuss another method, which is based on finding the exact solution of a non-linear extension of the wave equation \cite{eleuch2010new,eleuch2015localization}.

\section{Non-linear approximation to the wave equation equation in a random potential}
The 1D wave equation (or Schr\"{o}dinger equation with $\hbar=2m=1$) is given by
\be
[\partial_x^2+p(x)^2]\psi (x)=0,
\label{schroedinger}
\ee
with classical momentum 
\be
p(x)\equiv\partial_x P(x)=\sqrt{V(x)-E},
\ee
where we have defined $P(x)$ as the integrated momentum.

When looking for a solution of the form
\be
\psi(x)=e^{i(P+N)}=e^{i\int_{0}^{x}f(x')dx'},
\label{pplusn}
\ee
normalized at $x=0$, we obtain the following non-linear equation for $N(x)$
\be
i(\partial_xN)^2+2ip\partial_xN+\partial_xp+\partial_x^2 N=0,
\label{ERSapprox} 
\ee
which is difficult to solve \cite{eleuch2010new,eleuch2010analytical}. So instead, we can solve the related non-linear wave equation
\be
[\partial_x^2+p^2-[\psi^{-1}(-i\partial_x-p)\psi]^2]\psi=0, 
\label{schroedingerNL}
\ee
which leads to a linear equation in $N(x)$:
\be
2ip\partial_xN+\partial_xp+\partial_x^2 N=0
\label{ERS}
\ee
using equ. (\ref{pplusn}). This is equivalent to assuming $(\partial_xN)^2=0$ in equ. (\ref{ERSapprox}). The non-linear approximation corresponds to neglecting the difference between the classical and quantum momentum to second order: $((p+i\partial_x)\psi)^2\simeq 0$. The differential operator corresponding to equ. (\ref{ERS}) is then
\ba
H_N&=&2ip\partial_x+\partial_x^2\nonumber\\
&=&e^{-2iP}\partial_x(e^{2iP}\partial_x ),
\label{1D}
\ea
where we need to solve
\be
H_N N=-\partial_xp.
\ee
The solution can be obtained by integration, i.e.,
\be
\partial_x N(x)=-e^{-2iP(x)}\int^x e^{2iP(x')}k'_v(x')dx'\nonumber\\
\ee
to give
\be
f(x)=p(x)-e^{-2iP(x)}\int^x e^{2iP(x')}p'(x')dx'.\nonumber\\
\label{fplus}
\ee
The average over disorder can be performed, which gives
\be
\langle f(x) \rangle=\langle p(x) \rangle-\int_0^xe^{-2ik_0x'}c_p(x'),
\label{f}
\ee
where
\be
c_p(x)=\langle k'(0) e^{-2iP(x)}\rangle
\ee
is the correlation function of the disorder potential, where we defined the average momentum $k_0=\langle p(x)\rangle$), the variation from the mean ($k(x)=p(x)-k_0$) and its spatial derivative ($k'(x)$). To obtain (\ref{f}) one has to assume that the disorder is translationally uniform. The decay of the wave solution (Lyapounov exponent, $\lambda$ or inverse localization length) is then given by \cite{eleuch2015localization,hilke2017anderson}
\be \lambda=\Im \langle f(x) \rangle,\label{lambdaNL}\ee 
which can be rewritten as
\begin{equation}
\lambda = -2\Re\left\{ \int_{0}^{\infty }\left(
k_{0}C_{1}(x)+C_{2}\left( x\right) \right) e^{-2ik_{0}x}dx\right\},
\label{lambdaIntegral}
\end{equation}
where 
\begin{equation}
C_{n}(x)=\left\langle k(0)^{n}e^{-2iP(x)}\right\rangle.
\label{c(n)}
\end{equation}

This equation was shown to be valid for all disorder strengths for a number of disorder correlations, such as Gaussian \cite{eleuch2015localization}, speckle \cite{hilke2017anderson} and square well potentials \cite{eleuch2018ers}. In the limit of weak disorder the well known Born approximation is retrieved \cite{izrailev2012anomalous,eleuch2015localization}:
\be
\lambda_{\tilde{V}}=\frac{|\tilde{V}(2k_0)|^2}{8EN},
\ee
where $E=k_0^2$ and $\tilde{V}$ is the Fourier transform of the disorder potential. This weak disorder approximation can also be derived directly using Fermi's golden rule. 

\section{Arbitrary correlated potentials}

For arbitrary correlations, we obtain the correlators $C_n(x)$ as defined in equ. (\ref{c(n)}) by assuming that $k(x)$ are Gaussian variables (see appendix \ref{app}), where 

\begin{eqnarray}
c(x) &=&\langle k(0)k(x)\rangle  \notag \\
C(x) &=&\int_{0}^{x}c(y)dy  \notag \\
\mathscr{C}(x) &=&\int_{0}^{x}C(y)dy  \notag \\
C_{0}(x) &=&e^{-4\mathscr{C}(x)}  \notag \\
C_{1}(x) &=&-2iC(x)C_{0}(x)  \notag \\
C_{2}(x) &=&C_{0}(x)\left( c(0)-4C^{2}(x)\right).
\label{Cexpressions}
\end{eqnarray}

For numerical comparisons we want to be able to generate disorder potentials with arbitrary correlations, i.e., for a given correlation $c(x)$ we want to construct a disorder potential $k(x)$, which follows $\langle k(0)k(x)\rangle=c(x)$. In addition we will impose that $k(x)$ are Gaussian variables (only even correlators are non-zero). We start with an uncorrelated stochastic function $\Gamma_x$ with Gaussian variables, where $\langle \Gamma_x\Gamma_{x'}\rangle = \delta(x-x')$. The disorder potential is then given by the convolution
\begin{equation}
k(x)=\int g(x-x')\Gamma(x')dx',
\label{k(g)}
\end{equation}
where $g(x)$ can be expressed by the desired correlation $c(x)$ in the following way:
\ba
\langle k(x)k(x')\rangle&=&\int\int g(x-x_1)g(x'-x_2)\langle \Gamma_{x_1}\Gamma_{x_2}\rangle dx_1dx_2\notag\\
&=&\int g(x-x_1)g(x'-x_1) dx_1.
\ea
Hence,
\begin{equation}
c(x)=\langle k(0)k(x)\rangle=\int g(-x_1)g(x-x_1) dx_1, 
\end{equation}
which is simply the convolution $c(x)=(g*g)(x)$ for $g$ symmetric. Using the Fourier transform $\mathcal{F}$ and its inverse $\mathcal{F}^{-1}$, we find the following expression fro $g(x)$:
\be 
g(x)=\mathcal{F}^{-1}(\sqrt{\mathcal{F}[c(x)]}),
\label{g}
\ee
which allows us to construct a disorder potential with correlation $c(x)$ (see equ. (\ref{k(g)}). There is no unique method to construct the correlated potential using $g$. In fact, $k(x)$ can also be constructed using the following sum: 
\begin{equation} k(x)=\sum_i v_ig(x-x_i),
\label{k(v)}
\end{equation}
where $v_i$ and $x_i$ are independent random Gaussian variables. In this case we also have 
\be \langle k(0)k(x)\rangle\sim c(x)\ee. 

A similar inverse Fourier method was used for discrete random potential with arbitrary long range correlations \cite{makse1995novel}. Other methods that can generate correlated binary sequences use an iterative technique \cite{usatenko2014iterative}.

\begin{figure}[h]
	\centering
	\includegraphics[width=0.5\textwidth]{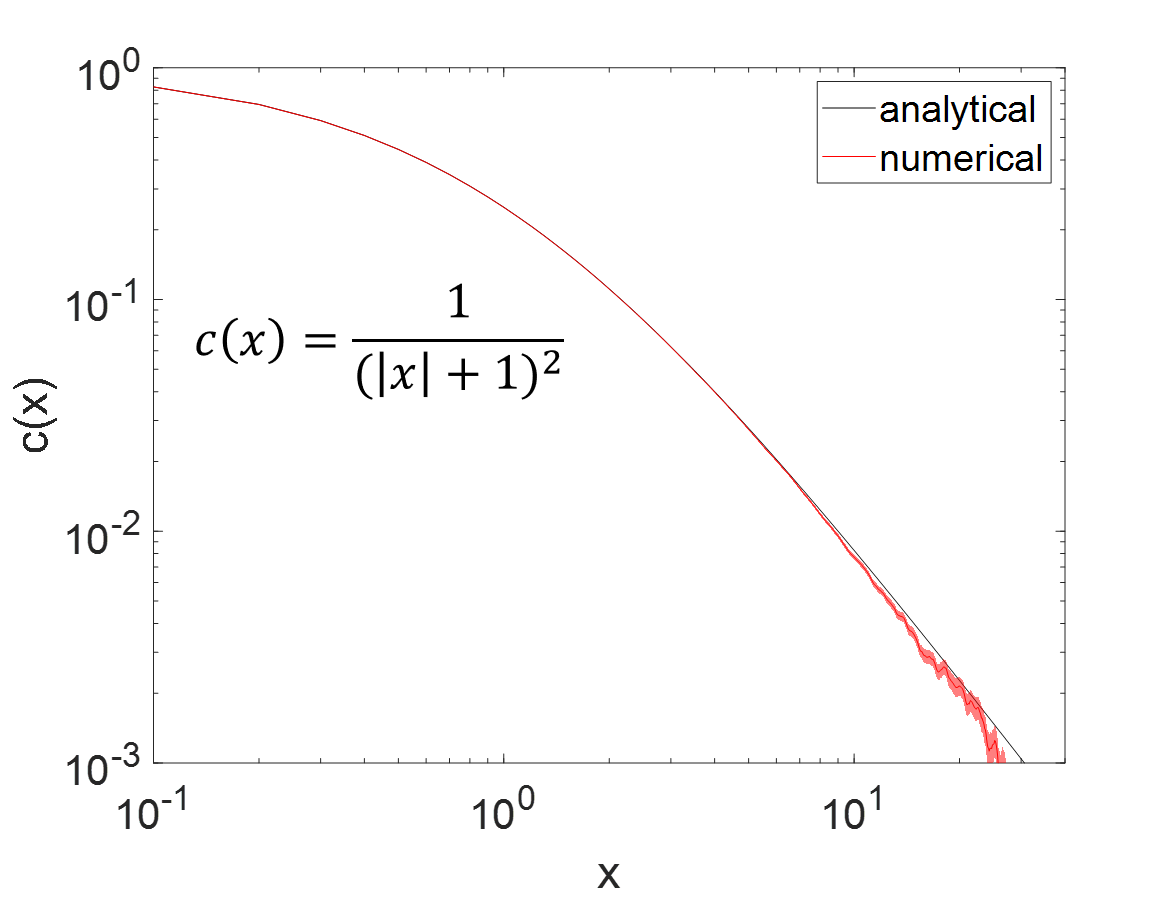}
	\caption[autocorrelation]{Autocorrelation $c(x)$ (in black) and the numerical evaluation of $\langle k(0)k(x) \rangle $ in red. The width of the red line is the numerical error of the mean for an averaging over 4000 configurations. The small systematic deviation at large $x$ is a finite size effect.}
	\label{correlation}
\end{figure}

\section{Localization for a quadratic decaying correlation function}

We illustrate the effectiveness of equs. (\ref{k(g)},\ref{g}) with an example of a correlation function, which is long-ranged and which decays quadratically with distance:
\begin{equation}
c\left( x\right)=\frac{\sigma^2}{\left( \left\vert x/a\right\vert +1\right) ^{2}},
\label{c(x)}
\end{equation}
with $a>0$ the correlation length and $\sigma^2=\langle k^2 \rangle$ the disorder strength. The numerically computed correlation function obtained using equ. (\ref{k(v)}) is shown in fig.  \ref{correlation} and it matches the desired correlation function $c(x)$. The same procedure can be applied to any choice of correlation function. This is illustrated by comparing the disorder potential generated using equ. \eqref{c(x)} (with $a=1$ and $\sigma=1$) and the Lorentzian correlation $(x^2+1)^{-1}$ using the same random sequence from equ. \eqref{k(v)} but a different generating function $g$.

\begin{figure}[h]
	\centering
	\begin{tabular}{c}
 \includegraphics[width=0.5\textwidth]{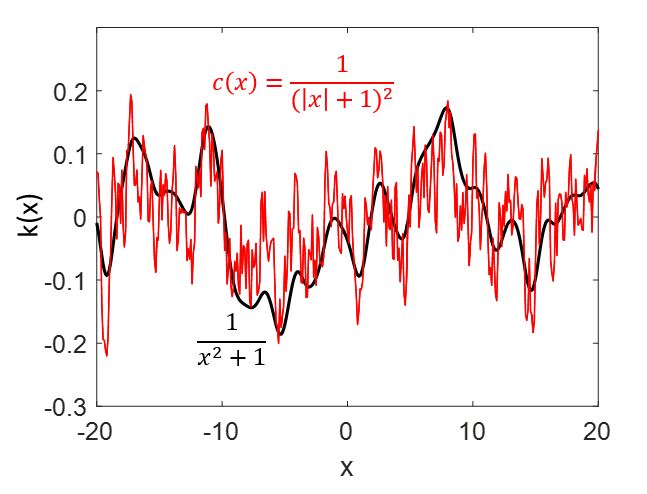} 
	\end{tabular}%
	\caption[autocorrelation]{Ccorrelated potentials $k(x)$ for autocorrelations $c(x)=\frac{1}{(|x|+1)^2}$ (in red) and $\frac{1}{x^2+1}$ (in black). }
	\label{V}
\end{figure}

\begin{figure}[h]
	\centering
		\includegraphics[width=0.5\textwidth]{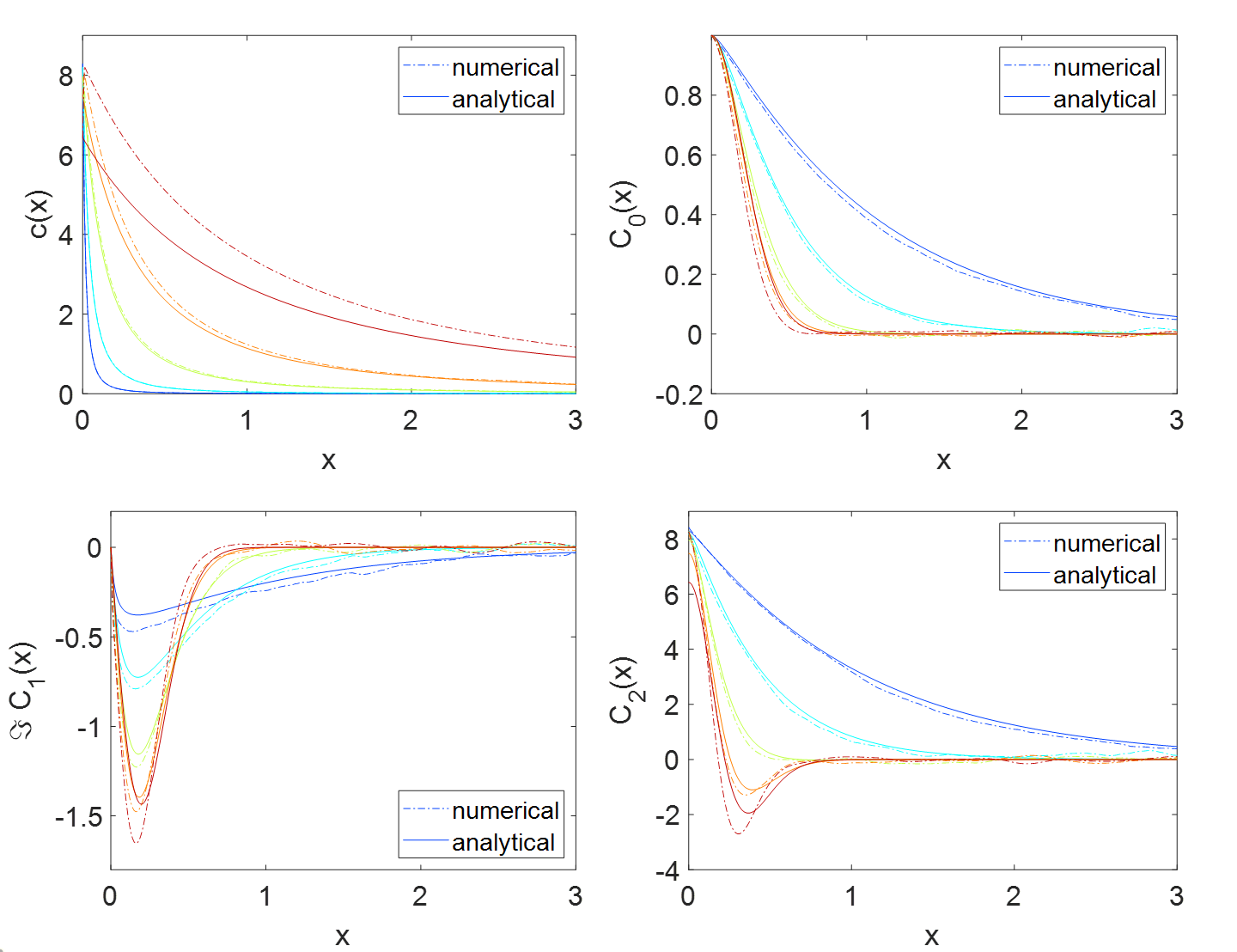} 
	\caption[cp]{The correlators $C_n(x)$, where the dashed lines are numerical simulations (averaged over 10000 configurations), while the solid lines are the analytical expressions from equs. \eqref{C0bis}. The different colors correspond to different values of the correlation length ($a$) from blue to red (0.03, 0.08, 0.2, 0.6, 1.8).}
	\label{cp1}
\end{figure}

We can compute explicitly the localization length associated to our choice of correlator $c(x)$ from equ. \eqref{c(x)} using relations \eqref{Cexpressions} and obtain for the various correlators (expressing only the case where $x$ is positive):
\begin{eqnarray}
C(x)&=&\frac{\sigma^2x}{x/a+1}\notag \\
\mathscr{C}(x)&=&\int_{0}^{x}dz\int_0^zdyc(y)=\sigma^2ax-\sigma^2a^2\ln \left[1+x/a\right]\notag \\
C_{0}(x)&=&e^{-4\sigma^2ax}\left(1+x/a\right) ^{4\sigma^2a^2}\notag\\
C_{1}(x)&=&-2i\sigma^2xe^{-4\sigma^2ax}\left(1+x/a\right) ^{4\sigma^2a^2-1}\notag\\
C_{2}(x)&=&\left[\sigma^2(x/a+1)^2-4\sigma^4x.^2\right]e^{-4\sigma^2ax}\left(1+x/a\right) ^{4\sigma^2a^2-2}.\notag\\
\label{C0bis}
\end{eqnarray}
The figure of the correlators \eqref{C0bis} are shown in fig. \ref{cp1}

An explicit expression for the Lyapounov exponent $\lambda $ can now be obtained by using \eqref{lambdaIntegral} (see appendix \ref{app2})

\be
\lambda=\frac{y+\Im{I_\lambda}}{a(1-y)},
\label{LyapounovAna}
\ee
where
\be
I_\lambda=\frac{[2aky+2ia^2(k^2-\sigma^2)]e^{\bar{y}}\,\Gamma_{1+y}(\bar{y})}{\bar{y}^{y+1}},
\ee
$y=4a^2\sigma^2$ and $\bar{y}=y+2iak$.

\begin{figure}[h]
	\centering
	\includegraphics[width=0.5\textwidth]{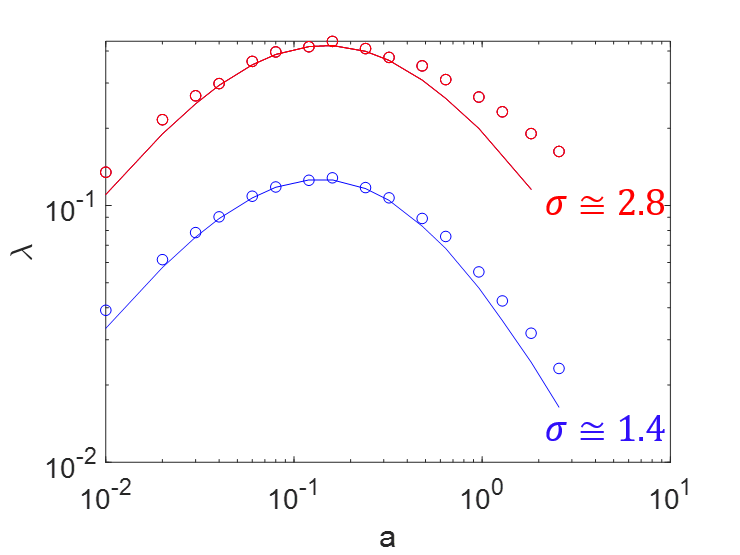} 
	\caption[cp]{The Lyapounov exponent ($\lambda$) obtained numerically (symbols) and expression \eqref{LyapounovAna} (solid lines) as a function of the correlation length $a$ ($E=20$).}
	\label{lyap}
\end{figure}

In fig. \ref{lyap} we show the comparison between the numerically evaluated $\lambda$ and expression \eqref{LyapounovAna} for the case of strong disorder. The disorder correlation is given by $c(x)$ (equ. \eqref{c(x)}). The overall agreement is very reasonable, in particular, when considering the wide range of correlation lengths $a$ and strong disorder. 

\section{Conclusion}

In summary, we review a method to calculate the localization length for a disordered potential with an arbitrary autocorrelation function, which uses the addition of a small non-linear term in the wave equation. The Lyapounov exponent is calculated by evaluating several correlators explicitly. For a comparison between theory and numerical simulations, we introduced a method that can generate a disorder potential with an arbitrary disorder correlation. We compared favorably the numerical results of a particular long ranged disorder potential, whose autocorrelation decays quadratically with distance, with our theory and find an excellent agreement. This method is quite general can be used to study other types of correlated potentials and is not restricted to weak disorder.

\appendix

\section{Appendix section}\label{app}

\subsection{Evaluating $C_0(x)$}

To evaluate the correlation functions $C_n(x)$ we consider $k(x)$ to be a random Gaussian variable with $\left\langle k(x)\right\rangle =0$. By definition, we have

\begin{eqnarray}
C_{0}(x) &=&\left\langle e^{-2i\int_{0}^{x}k(x^{\prime })dx^{\prime
}}\right\rangle \notag\\
&=&\overset{\infty }{\underset{n=0}{\sum }}\frac{1}{n!}\left\langle \left[
-2i\int_{0}^{x}k(x^{\prime })dx^{\prime }\right] ^{n}\right\rangle\notag \\
&=&\overset{\infty }{\underset{n=0}{\sum }}\frac{(-2i)^{n}}{n!}%
\int_{0}^{x}dx_{1}\cdots\int_{0}^{x}dx_{n}\left\langle
k(x_{1})\cdots k(x_{n})\right\rangle.\notag\\
\end{eqnarray}

For Gaussian random variables the odd number of correlators ($2n+1$) vanish
\begin{equation}
\left\langle k(x_{1})k\left( x_{2}\right)\cdots k(x_{2n+1})\right\rangle =0, 
\end{equation}
while for even number of correlators ($2n$) we have
\begin{eqnarray}
\left\langle k(x_{1})k\left( x_{2}\right)\cdots k(x_{2n})\right\rangle
&=&\left\langle k(x_{1})k\left( x_{2}\right)
\rangle\cdots\langle k(x_{n-1})k(x_{2n})\right\rangle  \notag \\
&+&\left\langle k(x_{1})k\left( x_{3}\right)
\rangle\cdots \langle k(x_{n-1})k(x_{2n})\right\rangle  \notag \\
&&\cdots  \notag \\
&+&\left\langle k(x_{1})k\left( x_{2n}\right)
\rangle\cdots \langle k(x_{2n-2})k(x_{2n-1})\right\rangle . \notag\\
\end{eqnarray}

Hence

\begin{eqnarray}
C_{0}(x) &=&\overset{\infty }{\underset{n=0}{\sum }}\frac{(-2i)^{2n}}{(2n)!}%
\int_{0}^{x}dx_{1}\cdots\int_{0}^{x}dx_{n}\left%
\langle k(x_{1})\cdots k(x_{2n})\right\rangle  \notag \\
&=&\overset{\infty }{\underset{n=0}{\sum }}\frac{\left( -2i\right)
	^{2n}\left( (2n-1)!!\right)}{(2n)!} \left[ \int_{0}^{x}\int_{0}^{x}\left\langle
k(x_{1})k(x_{2})\right\rangle dx_{1}dx_{2}\right] ^{n}  \notag \\
&=&\overset{\infty }{\underset{n=0}{\sum }}\frac{1}{(n)!}\left[
-2\int_{0}^{x}\int_{0}^{x}\left\langle k(x_{1})k(x_{2})\right\rangle
dx_{1}dx_{2}\right] ^{n}  \notag \\
&=&e^{-2\int_{0}^{x}\int_{0}^{x}\left\langle k(x_{1})k(x_{2})\right\rangle
	dx_{1}dx_{2}}.
\end{eqnarray}
This expression is similar to the result obtained in \cite{li2016quantum}. The argument $arg$ of the exponential can be evaluated as 
\begin{eqnarray}
arg&=&-2\int_{0}^{x}\int_{0}^{x}\left\langle k(x_{1})k(x_{2})\right\rangle
dx_{1}dx_{2}  \notag \\
&=&-2\int_{0}^{x}\int_{0}^{x}c(x_1-x_2)dx_{1}dx_{2}  \notag \\
&=&-2\int_{-x}^{x}c(z)(x-|z|)dz  \notag \\
&=&-4\int_{0}^{x}c(z)(x-z)dz \quad\mbox{ assuming $c(-z)=c(z)$}  \notag \\
&=&-4\int_{0}^x dy\int_{0}^y dz \, c(z)\notag\\
&\equiv & -4\int_{0}^x dy \, C(y)\notag\\
&\equiv & -4\mathscr{C}(x).
\end{eqnarray}
Hence,
\be C_0(x)=e^{-4\mathscr{C}(x)}.\ee

\subsection{Evaluating $C_1(x)$}
By definition, we have

\begin{equation}
C_{1}(x)=\left\langle k(0)e^{-2i\int_{0}^{x}k(x^{\prime })dx^{\prime
}}\right\rangle. 
\end{equation}
Using the relations $\langle ab^{n}\rangle =n\langle ab\rangle\langle
b^{n-1}\rangle$ for $a$ and $b$ two Gaussian variables, we have

\begin{eqnarray}
C_{1}(x) &=&\left\langle k(0)e^{-2i\int_{0}^{x}k(x^{\prime })dx^{\prime
}}\right\rangle  \notag \\
&=&\overset{\infty }{\underset{n=0}{\sum }}\frac{(-2i)^{n}}{(n)!}\left\langle k(0)
\left(\int_{0}^{x}dx'k(x')\right)^n\right\rangle  \notag \\
&=&\left\langle k(0)\right\rangle +\overset{\infty }{\underset{n=0}{\sum }}%
\left\{ \frac{2n+1}{(2n+1)!}\left\langle -2ik(0)\int_{0}^{x}k(x^{\prime
})dx^{\prime }\right\rangle \right.  \notag \\
&&\times \left. \left( \frac{2n!}{2^{n}n!}\right) \left\langle \left(
-2i\int_{0}^{x}k(x')dx'\right)^2\right\rangle
^{n}\right\}  \notag \\
&=& -2i\int_{0}^{x}\langle k(0)k(x^{\prime })\rangle dx^{\prime }\notag\\
&&\times\overset{\infty }{\underset{n=0}{\sum }}\frac{1}{n!}\left\langle
-2\int_{0}^{x}\int_{0}^{x}k(x_{1})k(x_{2})dx_{1}dx_{2}\right\rangle ^{n}
\notag \\
&=& -2i\int_{0}^{x}\langle k(0)k(x^{\prime })\rangle dx^{\prime } \times C_0(x)\notag \\
&=&-2iC(x)C_0(x),
\end{eqnarray}
where we have defined 
\be C(x)=\int_{0}^{x}\langle k(0)k(x^{\prime })\rangle dx^{\prime }.\ee

\subsection{Evaluating $C_2(x)$}
Using the relations $\langle a^{2}b^{n}\rangle =\langle a^{2}\rangle\langle b^{n}\rangle +n\langle ab\rangle \langle
ab^{n-1}\rangle$, $\langle a^{2}b^{n}\rangle=\left\langle a^{2}\right\rangle \left\langle b^{n}\right\rangle
+n(n-1)\left\langle ab\right\rangle ^{2}\left\langle b^{n-2}\right\rangle $ and $\left\langle b^{2n-1}\right\rangle=0$ for a and b Gaussian variables, we obtain 

\begin{eqnarray}
C_{2}(x) &=&\left\langle k^{2}(0)e^{-2i\int_{0}^{x}k(x^{\prime })dx^{\prime
}}\right\rangle  \notag \\
&=&\overset{\infty }{\underset{n=0}{\sum }}\frac{(-2i)^{n}}{n!}\left\langle k^2(0)
\left[\int_{0}^{x}dx' k(x')\right]^n\right\rangle  \notag \\
&=&\langle k^2(0)\rangle \overset{\infty }{\underset{n=0}{\sum }%
}\frac{(-2i)^{n}}{n!}\left\langle\left[
\int_{0}^{x}dx' k(x')\right]^n\right\rangle  \notag \\
&+&\overset{\infty }{\underset{n=2}{\sum }}\frac{(-2i)^{n}n(n-1)}{n!}\left%
\langle k(0)\int_{0}^{x}dx^{\prime }k(x^{\prime })\right\rangle ^{2}
\notag \\
&\times&\left\langle\left[
\int_{0}^{x}dx'k(x')\right]^{n-2}\right\rangle  \notag \\
&=&c(0)C_0(x)+\left\langle -2i\int_{0}^{x}k(0)k(x^{\prime })dx^{\prime }\right\rangle
^{2} \notag \\
&\times&\overset{\infty }{\underset{n=2}{\sum }}\frac{(-2i)^{n-2}}{(n-2)!}
\left\langle\left[
\int_{0}^{x}dx'k(x')\right]^{n-2}\right\rangle  \notag \\
&=&c(0)C_{0}(x)-4C^2(x)C_{0}(x).
\end{eqnarray}

\subsection{Lyapounov exponent}\label{app2}

The expression for $\lambda $ can be calculated as follows using equ. \eqref{lambdaIntegral}:

\begin{eqnarray}
\lambda & =& -2\Re\left\{ \int_{0}^{\infty }\left(
k_{0}C_{1}(x)+C_{2}\left( x\right) \right) e^{-2ik_{0}x}dx\right\}\notag\\
&=&-2ky\Im \left( \frac{e^{y+2iak}E_{-y}\left(y+2aik\right) }{y-1}\right)  \nonumber\\
&&-2a\Re \left( \frac{2\sigma^2+(k^2-\sigma^2)e^{y+2iak}E_{-y}\left(y+2aik\right) }{y-1}\right)   \nonumber\\
  &=&\frac{y+\Im \left\{ e^{y+2iak}(y+2iak)^{-y-1}\Gamma_{1+y}(y+2iak)[2aky+2ia^2(k^2-\sigma^2)]\right\}}{a(1-y)}, \notag \\
\label{LyapounovAna}
\end{eqnarray}
where the exponential integral is defined by

\begin{equation}
E_{\alpha }(z)=\int_{1}^{\infty }\frac{e^{-zt}}{t^{\alpha }}dt,
\end{equation}
using $E_\alpha(z)=z^{\alpha-1}\Gamma_{1-\alpha}(z)$ and taking $y=4a^2\sigma^2$.

\bibliographystyle{apsrev4-1}
\bibliography{CorrelationXXX}   

\end{document}